\documentclass[manuscript,nonacm]{acmart}
\makeatletter
\def\@ACM@checkaffil{
    \if@ACM@instpresent\else
    \ClassWarningNoLine{\@classname}{No institution present for an affiliation}%
    \fi
    \if@ACM@citypresent\else
    \ClassWarningNoLine{\@classname}{No city present for an affiliation}%
    \fi
    \if@ACM@countrypresent\else
        \ClassWarningNoLine{\@classname}{No country present for an affiliation}%
    \fi
}
\makeatother

\makeatletter
\let\@authorsaddresses\@empty
\makeatother
\AtBeginDocument{%
  }

\setcopyright{rightsretained}
\copyrightyear{2024}
\acmYear{2024}
\acmDOI{10.1145/3630106.3658899}

\acmConference[FAccT '24]{The 2024 ACM Conference on Fairness, Accountability, and Transparency}{June 3--6, 2024}{Rio de Janeiro, Brazil}

\acmBooktitle{The 2024 ACM Conference on Fairness, Accountability, and Transparency (FAccT '24), June 3--6, 2024, Rio de Janeiro, Brazil}

\acmISBN{979-8-4007-0450-5/24/06}

\usepackage{tikz,pgf,wrapfig, xcolor, enumitem}
\usetikzlibrary {arrows.meta}
\usetikzlibrary{backgrounds}
\definecolor{sepia}{HTML}{671800}
\definecolor{light-gray}{gray}{0.01}
\usepackage{xfrac}
\usepackage{stringstrings}

\begin{document}

\title[Algorithmic Pluralism]{Algorithmic Pluralism: A Structural Approach To Equal Opportunity}

\settopmatter{authorsperrow=1} 
\newcommand{\tsc}[1]{\textsuperscript{#1}} 
\author[S. Jain, V. Suriyakumar, K. Creel, A. Wilson]{Shomik Jain\tsc{1}, Vinith M. Suriyakumar\tsc{2}, Kathleen A. Creel\tsc{3}, \caselower{and} Ashia C. Wilson\tsc{1,2}}

\affiliation{\qquad \qquad \qquad \qquad \qquad \qquad \qquad \qquad\qquad \qquad \qquad  \qquad
\hspace{-5pt}  {\small \tsc{1} Institute for Data, Systems, and Society, MIT \quad \tsc{2} Department of Electrical Engineering and Computer Science, MIT{\color{white}\country{MIT}}\\
\tsc{3} Department of Philosophy and Religion and Khoury College of Computer Sciences, Northeastern University
}}

\begin{abstract}
We present a structural approach toward achieving equal opportunity in systems of algorithmic decision-making called \textit{algorithmic pluralism}. Algorithmic pluralism describes a state of affairs in which no set of algorithms severely limits access to opportunity, allowing individuals the freedom to pursue a diverse range of life paths. To argue for algorithmic pluralism, we adopt Joseph Fishkin’s theory of bottlenecks, which focuses on the structure of decision-points that determine how opportunities are allocated. The theory contends that each decision-point or ``bottleneck’’ limits access to opportunities with some degree of severity and legitimacy. We extend Fishkin's structural viewpoint and use it to reframe existing systemic concerns about equal opportunity in algorithmic decision-making, such as patterned inequality and algorithmic monoculture. In proposing algorithmic pluralism, we argue for the urgent priority of alleviating severe bottlenecks in algorithmic decision-making. We contend that there must be a \textit{pluralism of opportunity} available to many different individuals in order to promote equal opportunity in a systemic way. We further show how this framework has several implications for system design and regulation through current debates about equal opportunity in algorithmic hiring. 
\end{abstract}




\maketitle

\section{Introduction}
Almost all principles of fair decision-making derive their moral force from the principle of equality: the idea that all persons have equal moral worth and therefore should be treated equally. Since the general principle of equality could give rise to many specific types of equality, theorists of equality often debate what in particular should be equalized, whether it be opportunity, welfare, esteem, material goods, or capabilities~\citep{SenEqualityOfWhat, Anderson1999, rawls2004theory}. This debate about what to equalize extends to the algorithmic fairness literature, with proponents of demographic parity arguing for equalizing acceptance rates~\citep{calders2009building, zliobaite2015relation}, proponents of equalized odds advocating for equalizing the likelihood of particular decision outcomes~\citep{hardt2016equality}, and proponents of calibration arguing for equalizing the meaning of model predictions \citep{pleiss2017calibration, hebert-johnsonMulticalibration}. Amidst this disagreement, \textit{equality of opportunity}, the idea that everyone should have the same opportunity to succeed, enjoys a broad consensus. Equal opportunity posits a world in which circumstances of birth and parentage do not fully determine one's life outcomes. It is a value that is singled out in several national founding documents and is regularly invoked by advocates of radically different political and social agendas~\citep{fishkin2014oligarchy}. The principle of equal opportunity also underlies social programs with widespread support around the world, such as public education, disaster relief, and food assistance.   

What does equality of opportunity mean in the context of algorithmic decision-making? Formal views of equal opportunity often contend that all decision subjects should have a chance at a positive outcome and that their likelihood of receiving that outcome accords with their merit or desert. The fairness metric of equalized odds, for instance, checks if the predictor and protected attributes (e.g. race or gender) are independent conditional on the outcome~\citep{hardt2016equality}. However, a growing number of scholars~\citep{Green2022, eaamo2022equal, kasirzadeh2022feminist, hoffmann2019fairness, selbst2019fairness} have criticized formal algorithmic fairness interventions that focus only on properties of model predictions and evaluate each decision-instance separately. They instead advocate for more \textit{structural views of equal opportunity} that take into account repeated encounters with decision-making systems as well as broad patterns of life chances and pathways~\citep{Green2022}. For example, a structural critique put forth by Benjamin Eidelson~\citep{eidelson2021patterned} questions systems of decision-making that consistently penalize those with already poor outcomes and reward those with good outcomes. Such a system, he argues, tends to create a ``patterned inequality'' in which existing disadvantage causes further disadvantage, compromising the possibility of equal opportunity and leading to a variety of social-ills and overall suffering. Using algorithms to support decision-making is likely to reinforce patterns of inequality due to their ability to recognize and precisely exploit existing social patterns. Another related concern about how opportunities are structured critiques a state of affairs in which the same individuals or groups of people are on the losing side of a large number of social allocations -- a phenomenon which Ifeoma Ajunwa calls ``(algorithmic) blackballing''~\citep{Ajunwa2021}, Kathleen Creel and Deborah Hellman call ``systemic exclusion''~\citep{creel_hellman_2022}, and Rishi Bommasani et al. call ``outcome homogenization''~\citep{bommasani_outcome_homogenization}. Data-driven decision-making could exacerbate these concerns due to a growing ``algorithmic monoculture''~\cite{kleinberg2021algorithmic} in which decision-makers that collectively dominate a sector rely on similar datasets and/or models and subsequently arrive at the same decisions.

In this work, we respond to the threat that patterned inequality and algorithmic monoculture pose to equal opportunity by arguing for the value of \textit{algorithmic pluralism}. Algorithmic pluralism is the state of affairs in which the decision-making algorithms that structure opportunity are meaningfully pluralistic, which is to say that their decisions result in a plurality of paths to different outcomes, in some cases because the decisions are made on the basis of different features and values. To argue for algorithmic pluralism, we adopt Joseph Fishkin's structural approach towards equal opportunity~\citep{fishkin2014bottlenecks} and extend it to algorithmic decision-making. Fishkin highlights the importance of \textit{bottlenecks}, which are decision points or ``narrow places'' in the structure of opportunity. Bottlenecks are the mechanisms through which opportunities are created, distributed, or destroyed. In these real-world networks of decisions, bottlenecks often chain together so that the output of one decision point is an input to the next. For example, resume-screening algorithms operate as a bottleneck for many job  opportunities. Each bottleneck can be evaluated in terms of its ``severity,'' or degree to which it constrains opportunities, and its ``legitimacy,'' or its justification in relation to how it allocates opportunities. Fishkin highlights that systems of decision-making with severe bottlenecks should often be considered an infringement of equal opportunity, regardless of their legitimacy. Subsequently, alleviating severity by expanding the number of pathways to opportunity should be an urgent priority. 

We extend this idea of \textit{opportunity pluralism}~\citep{fishkin2014bottlenecks} to systems of algorithmic decision-making, resulting in our argument for \textit{algorithmic pluralism}. Specifically, we make the following contributions:
\begin{itemize}
    \item We review classic tensions in the equality of opportunity literature to motivate Fishkin's idea of opportunity pluralism (Section~\ref{sec:equalOpportunity}).
    
    \item We present Fishkin's theory of bottlenecks and use the framework to center several structural concerns about equal opportunity raised by a variety of scholars who focus on algorithms and issues of justice (Section~\ref{sec:bottleneck}).
    
    \item We propose algorithmic pluralism, highlighting the importance of alleviating severe bottlenecks in algorithmic-decision-making and the need to make many kinds of opportunity available to many different individuals (Section~\ref{sec:pluralism}).

    \item We motivate and illustrate the value of algorithmic pluralism by applying it to current debates about equal opportunity in algorithmic hiring (Section~\ref{sec:caseStudies}).
    
    \item We end our discussion with implications for regulation and design (Section~\ref{sec:implications}).
  
\end{itemize}

\section{Equal Opportunity and Its Limitations}\label{sec:equalOpportunity}
In order to motivate Fishkin's idea of opportunity pluralism, we will briefly explore classic tensions in the equality of opportunity literature that highlight the importance of considering both access to opportunity and the structure of opportunity itself. We begin by sketching a classic thought experiment proposed by Bernard Williams~\citep{williams1962warrior} and discussed by Fishkin~\citep{fishkin2014bottlenecks}. The thought experiment asks us to imagine a society in which there are only two social roles: warriors and non-warriors. In this society, warriors are highly valued. They receive all the material and non-material goods the society has, such as superior lodging, food, salary, opportunity for advancement, and social esteem. Since there are a fixed number of positions in the army, not all members of the society can become warriors. A meritocratic process is established to determine who can become a warrior: at the age of sixteen, all members of the society participate in a multi-day competition that is intended to measure characteristics such as athletic ability, endurance, bravery, and battlefield strategy.  

Imagine that an algorithm is created that uses the results of the competition to determine who should become a warrior. Under one interpretation of equal opportunity, a fair algorithm is one that gives all members of the society an equal opportunity to become a warrior by choosing warriors based only on features relevant to their success on the battlefield. However, is this algorithm truly fair? After all, children of warrior parents have a significant advantage. Since their parents are warriors themselves, they are better able to train their children in the martial arts. And because their parents are the wealthiest and most esteemed members of society, their children are more likely to be better fed, healthier, stronger, and more confident -- all traits that are reflected in the differential success rate of children from warrior and non-warrior parents. 
The children of warrior parents also benefit from additional experiences and resources, such as family wealth and access to better education, that better prepare them for the warrior test. 

These \textit{developmental opportunities} make it impossible for warrior and non-warrior children to have a fair contest. While non-warrior children are not explicitly excluded because of their caste, the data collected on them reflects their lack of developmental opportunities. For this reason, selecting warriors based on who will perform best without equalizing their childhood opportunities in any way will tend to disadvantage non-warrior children. As Fishkin explains, ``if success breeds success, and we reinforce achievement with new and richer developmental opportunities, then the project of equalizing opportunities comes squarely into conflict with rewarding performance. In that case, the very earliest developmental opportunities, which precede any meaningful performances worth rewarding, begin to take on an outsized significance''~\citep[p5]{fishkin2014bottlenecks}. We extend this thought experiment to highlight additional problems with formal translations of equal opportunity in the context of algorithmic decision-making. 

\setlength{\parindent}{0cm}
\paragraph{\textbf{The problem of a starting gate}}\quad{}\\
Given that childhood disadvantage becomes disadvantageous~\citep{eidelson2021patterned}, making later meritocratic contests unfair, one of the most important principles of equal opportunity has been to equalize developmental opportunities by providing all children (and later, all people) with equal access to education, nutrition, and other building blocks of development.  This approach is sometimes called equalizing the \textit{starting gate} in the race of life, ensuring that all the competitors show up with equal opportunity to prepare with the hopes that decision-making on the basis of merit after that point will be fair as a result~\citep{rawls2004theory}. Since all agree that it is impossible to fully equalize the starting gate -- even with equal material opportunity, some parents may be kinder and more supportive than others -- some scholars have attempted to re-create an equal starting gate through statistical methods~\citep{hardt2022backward}.

\paragraph{\textbf{The problem of isolating merit}}\quad{}\\
Allocating goods on the basis of later merit without equalizing developmental opportunities undermines the normative rationale for allocating goods on the basis of merit at all. If one justification for allocating goods on the basis of merit is that it rewards hard work and dedication, a profoundly unequal playing field makes it impossible to isolate those qualities from advantageous circumstances of birth and life experiences. As Fishkin notes, ``everything we are and everything we do is the product of layer upon layer of interaction between person and environment -- between our selves, our efforts, and our opportunities -- that in a sedimentary way, over time, build each of us into the person we become''~\citep[p8]{fishkin2014bottlenecks}. Because of this, any real-world data and features collected to become the basis of algorithmic decision-making will inherently combine aspects of hard work, innate talent, and developmental opportunities in ways that are difficult to distinguish. 

\paragraph{\textbf{The problem of metrics and what is measured}}\quad{}\\
Another challenge to the fair implementation of equal opportunity is that any test established to measure merit is unlikely to perfectly measure real-world performance. Warrior children may score higher on any test of athletic ability, for example, because of they have received special coaching that improves their scores on the test more. However, this coaching may improve test scores more than it improves real-world battlefield performance. In general, given it is impossible  to measure merit precisely, proxy metrics will be used which will inevitably unfairly limit the opportunity of some. 

\paragraph{\textbf{The problem of zero-sum thinking}}\quad{}\\
Focusing on specific decision points, such as the test of sixteen year-old warriors, also needlessly reduces considerations of equal opportunity to zero-sum thinking. For example, if we choose to give compensatory bonus points to non-warrior children, some warrior children may object at no longer being selected. This distributive view forces zero-sum trade-offs between mathematically incompatible notions of fairness, magnifying the stakes of choosing between these definitions~\citep{kleinberg2016inherent}. As Green argues, restricting analysis to specific decision points also cannot fully ``account for the inequalities that often surround those decision points [and] is therefore prone to reproducing existing patterns of injustice''~\citep{Green2022}. By expanding our view to the broader structure of opportunities, we can better account for the important contexts in which decisions are made and consider interventions that diminish or even eliminate the zero-sum framing. \\

The warrior society thought experiment illustrates that focusing only on fairness metrics takes an overly narrow view of the opportunity structure. As Dewey astutely points out, ``the way in which [a] problem is conceived decides what specific suggestions are entertained and which are dismissed''~\citep{dewey1938logic}. Formal fairness methods often reduce considerations to relative advantages or disadvantages at a particular decision point. This type of single-axis thinking comes at the expense of attention to what produces the systematic benefits or privileges in the first place~\citep{hoffmann2019fairness}. In doing so, formal approaches do not account for the fact that having a plurality of opportunities (and not just equal opportunities) is a prerequisite for the accompanying virtues of freedom and the ability to shape our lives that make equal opportunity an attractive value in the first place.  
\setlength{\parindent}{.4cm}

\section{Bottleneck Theory: A Structural View of Equal Opportunity}\label{sec:bottleneck}
So far we have seen that establishing a fair starting gate and fair metrics for performance are \textit{necessary} for equal opportunity in a strict meritocratic society. Next, we follow Fishkin in arguing that they are not \textit{sufficient}.

\subsection{Opportunity Pluralism}
\label{subsec:opportunitypluralism}
The warrior society thought experiment reveals that equalizing developmental opportunity alone is insufficient for maintaining the virtues associated with equal opportunity. Imagine we can fully equalize developmental opportunities by placing all children into warrior skill academies from the moment of birth. While those not selected to become warriors now have equal developmental opportunities, they still may have a broader complaint about the \textit{structure of opportunity} in their society: that the bottleneck through which they must pass to succeed is too narrow. In Fishkin's terms, a society whose pathway to success is \textit{only} open to those with the same set of natural talents (e.g. javelin throwing, martial courage, and battlefield tactics) has established an overly ``severe bottleneck'' on opportunity ~\cite{fishkin2014bottlenecks}.

Such a severe narrowing of life paths compromises the notion that equal opportunity is a principle which offers the freedom to shape our lives. That is, it is not enough that opportunities be equal. For example, in the warrior society, children with intellectual capacities should have the opportunity to become scribes or advisors and children with creative abilities and dexterity should have the opportunity to become craftspeople. Given that natural talents and inclinations are not chosen, akin to the lack of choice in one's parents, a society that provides equal opportunity should provide many ways in which natural talents can lead to success. There must be a pluralism of opportunities, not merely equal access to one opportunity. Even more broadly, there must be multiple ways of assessing capacities for opportunities. Having only one test such as in the warrior society establishes a severe bottleneck centering on the method of assessment. Its flaws, weaknesses, and oversights become magnified, meaning that even a child who could be a talented warrior might have their skills overlooked if they are not precisely the skills measured by the test. In addition to a plurality of opportunities, there must be a plurality of assessments. These considerations lead Fishkin to propose four central principles of \textit{opportunity pluralism}~\citep{fishkin2014bottlenecks}:

\begin{enumerate}
    \item There should be plurality of values and goals.
    \item As many as possible of the valued goods should be less positional and the valued roles less competitive.
    \item As far as possible, there should be a plurality of paths leading to these different valued goods and roles, without bottlenecks constraining people’s ability to pursue those paths. 
    \item There should be a plurality of sources of authority regarding the elements described in the other principles. 
\end{enumerate}

\subsection{Bottlenecks}
Fishkin's opportunity pluralism centers on alleviating bottlenecks, or ``narrow place[s] in the opportunity structure''. A bottleneck is a crucial decision-point that affects how future opportunities are created, distributed, and controlled~\citep[p1]{fishkin2014bottlenecks}. Examples of bottlenecks include decision-points that allow one to gain credentials or develop skills that are required to pursue certain opportunities. Bottlenecks are often chained together given that in systems of decision-making, the outcomes of upstream decisions usually factor into downstream decisions. For instance, the college admissions process constitutes a significant bottleneck for many employment opportunities and life paths, but is also in itself controlled by the bottlenecks of standardized testing and access to extra-curricular activities. As algorithms are increasingly used to inform or make decisions about our lives, they can become bottlenecks that control access to many real-world opportunities such as employment, housing, and education. Legibility to resume screening algorithms is a bottleneck on the path to employment; approval by tenant screening algorithms is a bottleneck on the path to housing; and acceptance by college admissions algorithms may soon be a bottleneck on the path to certain elite institutions. In his framework for opportunity pluralism, Fishkin proposes the qualitative measures of \textit{severity} and \textit{legitimacy} to assess the impact of each bottleneck on the broader opportunity structure. We show how these measures map to several predominant concerns about algorithmic decision-making and its potential to infringe upon structural access to opportunity.

\begin{figure}
\begin{tikzpicture}[thick,scale=0.5,
    show background rectangle, 
    show background rectangle, 
    background rectangle/.style={fill=black!0},
    box/.style={draw, font=\itshape}
]]
\tikzstyle{myarrows}=[line width=1.8mm, ->, >=stealth, opacity=0.5]

        \node[above] at (5,8.7) {\small Severe};
        \node[above] at (5,8) {\footnotesize (Pervasive \& Strict) };
        \node[above] at (0,4) {\small Legitimate}; 
        \node[above] at (10,4) {\small Arbitrary}; 
        \node[above] at (5,-1) {\small Mild}; 
        \draw[draw=black,thick,<->] (5, 7.5) to (5, 0);
        \draw[draw=black,thick,<->] (0, 3.75) to (10, 3.75);
        \definecolor{tempcolor}{HTML}{556B2F}
        
    \end{tikzpicture}
    \Description[]{(Fig. 6 in~\citep[p167]{fishkin2014bottlenecks}) Fishkin's framework for classifying bottlenecks. He proposes mapping the structural impact of bottlenecks along the axes of severity and legitimacy.}
    \caption{(Fig. 6 in~\citep[p167]{fishkin2014bottlenecks}) Fishkin's framework for classifying bottlenecks. He proposes mapping the structural impact of bottlenecks along the axes of severity and legitimacy.}
    \label{fig:axes}
    \end{figure}
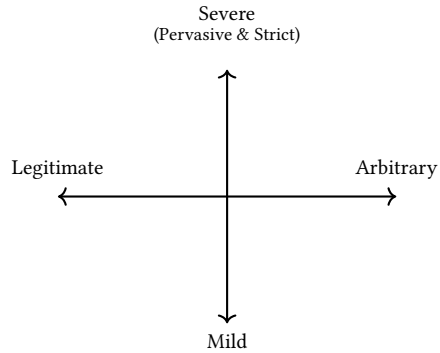

\subsection{Bottleneck Severity}

The severity of a bottleneck is the degree to which it constrains opportunities. Severity is a combination of two factors: pervasiveness and strictness~\citep[p164]{fishkin2014bottlenecks}. A bottleneck's \textit{pervasiveness} is the range of people and opportunities it affects. For example, the warrior selection algorithm in the warrior society has maximum pervasiveness because it controls everyone's access to all the valued goods in the society. \textit{Strictness} refers to how directly a bottleneck controls opportunities, or in other words, whether it is an absolute bar, strong preference, or weak preference. The warrior selection algorithm also has maximum strictness because it is an absolute bar: no one can become a warrior unless they are selected by the algorithm. While the warrior society represents an extreme case, many severe bottlenecks naturally arise in systems of data-driven decision-making, as illustrated by the concerns expressed in the ideas of {\em patterned inequality} and {\em algorithmic monoculture}. 

\setlength{\parindent}{0cm}
\paragraph{\textbf{The concern about patterned inequality is a concern about severe bottlenecks}}\quad \\
Patterned inequality refers to two observations made by Benjamin Eidelson: (1) real-world inequalities in status, resources, and opportunities are often patterned in terms of certain discernible, socially salient features; (2) allocating future outcomes and resources based on an individual's likelihood of success will predictably reproduce and aggravate patterns of this kind~\citep{eidelson2021patterned}. The problem of developmental opportunities makes it inevitable that some patterns of inequality will exist. Once an observable trait, perhaps implicit or arbitrary in itself, is correlated with less visible attributes relevant for deciding opportunities, decision-makers will rationally tend to use that trait as a proxy for allocating future outcomes (c.f. the problem of isolating merit). For example, consider the strong correlation between credit score and job performance, which led to its controversial use as a proxy for hiring decisions~\citep{ballance2020creditScore}. Given algorithms are developed to identify these patterns in a more explicit way, the concern is that allocating opportunities using them ``will tend to reproduce existing patterns in inequality and cement the matrix of stereotypes and social meanings that both cause and result from those patterns''~\citep{eidelson2021patterned}. This stronger tendency towards the comparatively privileged creates the possibility of stricter bottlenecks, while the denial of even more opportunities to members of worse-off groups represents more pervasive bottlenecks. 



\setlength{\parindent}{0cm}
\paragraph{\textbf{The concern about algorithmic monoculture is a concern about severe bottlenecks}}\quad \\
Algorithmic monoculture~\citep{kleinberg2021algorithmic} occurs when multiple decision-makers controlling access to a large quantity of valued goods rely on the same or similar datasets and/or models. As \citet{bommasani_outcome_homogenization} discuss, algorithmic monoculture can lead to a pattern of homogeneous outcomes in which the same people are subject to consistent errors or negative outcomes. \citet{creel_hellman_2022} have argued that at scale, this ``systemic  exclusion'' is unfair to decision-subjects even if overall accuracy is high. Algorithmic monocultures are increasingly common in high-stakes screening decisions across many domains, including in employment, healthcare, lending, and criminal justice~\citep{oneil2017weapons}. One reason why monoculture occurs is that few decision-makers have the resources to build their own automated systems. For example, over 60\% of Fortune 100 companies and 8 of the top 10 largest U.S. federal agencies use the same resume-screening service for hiring~\citep{hirevueArticle}. In other domains, the data collection itself may be expensive or unfeasible, such as in healthcare algorithms which disproportionately use data from a handful of large hospitals in only three states: California, Massachusetts, and New York~\citep{creel_hellman_2022}. Furthermore, the recent rise of pre-trained foundation models, which can be adapted to a wide range of downstream tasks, can also contribute to algorithmic monocultures and increase the likelihood of standardized errors~\citep{bommasani2021opportunities, toups2023ecosystem}.  
\setlength{\parindent}{.5cm}

The standardization of decision-making processes could create stricter bottlenecks because certain features will be stronger preferences for the opportunities at hand. In ranking problems like comparing job candidates, sharing an algorithm across multiple decision-makers can also reduce the overall quality of decisions, even if the algorithm is more accurate for any one decision-maker in isolation~\citep{kleinberg2021algorithmic, peng2023monoculture}. Moreover, if similar algorithms are uniformly applied across wide swathes of a single domain, outcomes could become homogenized resulting the systematic exclusion of groups and individuals from opportunities of great consequence~\citep{creel_hellman_2022, bommasani_outcome_homogenization}. Thus outcome homogenization constitutes a pervasive bottleneck for those that repeatedly receive undesirable decisions because they may find themselves locked out of many paths in the opportunity structure.

\subsection{Bottleneck Legitimacy}

Other structural concerns about equal opportunity center on a model's \textit{legitimacy}, or its justification in relation to how opportunities are allocated~\citep[p160]{fishkin2014bottlenecks}. The justification for an algorithm is often argued on two grounds: mathematical legitimacy and social legitimacy. \textit{Mathematical legitimacy} is simply the algorithm's accuracy with respect to a particular problem formulation. As we saw in the warrior society, however, a high accuracy alone does not fully justify the use of an algorithm to control real-world outcomes. The other necessary aspect to consider is an algorithm's \textit{social legitimacy}: what is the rationale for using a particular problem formulation (e.g. choosing a specific model, features, or training objective) to constrain the specific opportunity at hand? While many scholars have raised concerns about legitimacy of algorithmic decision-making procedures~\citep{green_realism, hoffmann2019fairness, kasirzadeh2022feminist}, we highlight two in particular that direct their legitimacy critique at the way opportunities are structured. 

\setlength{\parindent}{0cm}
\paragraph{\textbf{The concern about compounding injustice is a concern about illegitimate bottlenecks.}}\quad \\
Compounding injustice refers to Deborah Hellman's critique of decisions that take a prior wrong that a person has suffered, or its effects, as a reason for allocating future opportunities in a way that makes them still worse off~\citep{hellman2018compounding}. In many cases, adverse predictions by algorithms may involve the use of features that relate to prior victimization. For example, a prominent healthcare algorithm selecting patients for high-risk care programs used prior health costs as a proxy for future health needs~\citep{obermeyer2019proxy}. This ignored the well-documented injustices of the U.S. healthcare system that cause less money to be spent on black patients who have the same level of need as equally sick white patients. By denying future care on the basis of prior health costs, the algorithm ``compounded the initial injustice'' of unequal access to care. From the perspective of bottleneck theory, Hellman's formulation of compounding injustice and Eidelson's patterned inequality both surface similar concerns about the structure of opportunities; however, Hellman's argument raises the illegitimacy of the initial condition as central to moral considerations, whereas Eidelson centers the severity of unequal patterns of opportunity as sufficient to warrant moral concern. 

\setlength{\parindent}{0cm}
\paragraph{\textbf{The concern about relational harms is a concern about illegitimate bottlenecks.}}\quad \\
Relational harms refer to when algorithms fail to account for the power relations, social dynamics, and structural contexts that surround systems of decision-making in the real-world~\citep{green_realism, Green2022}. In particular, this involves scrutinizing what institutions, values, and norms cause social and material disparities in opportunities~\citep{Anderson1999, minow1990making, green_realism}. Explicitly considering the potential for relational harms can help justify algorithmic design choices and problem formulations. Whether or not certain input features and output predictions are legitimate, for instance, can depend on if they reduce dignitary and material disparities that reflect social hierarchies~\citep{Green2022}. For example, arguments against the legitimacy of COMPAS highlight that predictions of recidivism fail to account for the racial biases in policing that make re-arrest rates higher for minority populations~\citep{corbett2016COMPAS}. In addition, how much a decision-maker values mitigating various relational harms can also inform the choice of a training objective~\citep{davis2021reparation}. Companies making hiring decisions may have different values such as hiring from the local community, creating more diverse teams, or addressing past discrimination. Training an algorithm based on who has been hired in the past may undervalue the contributions that underrepresented applicants bring to a company. Ultimately, considerations of these broader contexts and how they affect an algorithm's social legitimacy will depend on one's choice of worldview~\cite{friedler2021possibility}. This may differ from the perspective of various stakeholders, such as those making decisions versus those seeking opportunities. 
\setlength{\parindent}{.4cm}

\section{Algorithmic Pluralism}\label{sec:pluralism}
In this section, we draw on the concept of opportunity pluralism to define \textit{algorithmic pluralism}. We first argue that alleviating severity is the most important problem to address in order to make opportunities more equal. The previous section showed that Fishkin's structural view of equal opportunity requires making bottlenecks both less severe and more legitimate. But as many including Fishkin have argued, severity alone can be sufficient to warrant intervention even if a bottleneck is legitimate. We then define algorithmic pluralism by considering different ways in which an algorithmic system could be pluralistic. Ultimately, we conclude that of these different kinds of pluralism, the most important is that algorithms allow individuals with diverse characteristics (skills, identities, life paths, etc.) to have a variety of chances to gain access to valued opportunities and goods.  

\subsection{Legitimacy Matters, but Severity Takes Precedent}

The importance of ensuring legitimacy in algorithmic decisions cannot be overstated. Legitimacy aims to ensure that decisions are justified in relation to how they allocate opportunities. An algorithm can achieve legitimacy by being produced in accordance with legitimate process, such as a democratic procedure, or by its alignment with fundamental democratic values such as accuracy, equality, fairness, and justice~\citep{costanza2018design, kasy2021fairness, Holm2023}. The algorithmic fairness community has often relied on formal interventions to increase legitimacy that center on some metric of ``fairness''. Recent works rightfully also encourage the community to consider the legitimacy of the broader opportunity structure in addition to the fairness of the algorithm itself~\citep{Green2022, kasirzadeh2022feminist, hoffmann2019fairness}. The convergence of this work suggests that an algorithm that lacks mathematical or social legitimacy\footnote{There exists significant political disagreement about what makes a decision legitimate. As Ben Green points out, resolving the disagreement involves ``grappling with contested notions of what types of inequalities are unjust and what evidence constitutes sufficient proof of social hierarchy''~\citep{Green2022}. Although we consider this grappling to be important and necessary work, for the purpose of this paper we will rely on the concepts of ``mathematical legitimacy'' and ``social legitimacy'', introduced in the previous section. However, we believe that our arguments would hold were other notions of legitimacy to be substituted. For discussions of algorithmic legitimacy, see \citep{Wang2023, Waldman2022, Holm2023}.} is in need of intervention. If an algorithm or its decision outcomes are not legitimate, then those subject to the algorithm have the right to demand that it become legitimate, or that it cease to have the power to affect their lives. Thus legitimacy alone can be a sufficient condition for intervention. 

We can sometimes determine whether an individual decision or a decision-making system is legitimate without evaluating the broader structure of opportunity in which the decision takes place. Severity, by contrast, can be observed solely by evaluating the influence of decision-makers on the ultimate pattern of outcomes for decision subjects. Fishkin concludes that: ``all this leaves us with complex problems of prioritization. In a world of myriad bottlenecks, we need to decide which ones to devote our efforts and scare resources to ameliorating. The question of how important it is to loosen any given bottleneck turns in significant part on how severe the bottleneck is''~\citep[p186]{fishkin2014bottlenecks}. In other words, the severity of decisions on the structure of opportunities -- especially (but not only) when illegitimate -- should take precedent. For systems of algorithmic decision-making, we similarly contend that severity warrants intervention on its own, as other works\footnote{As Fishkin notes, a bottleneck that is legitimate and severe is deserving of more scrutiny and moral concern because of its weight on the opportunity structure than one that is illegitimate but not severe. Importantly, many works advocating views beyond formal algorithmic fairness (e.g. Green~\cite{Green2022}) fail to make a distinction between prioritizing legitimacy and severity concerns.} have argued using different frameworks~\citep{eidelson2021patterned, bommasani_outcome_homogenization, creel_hellman_2022}. Eidelson’s concern about patterned inequality is a concern about severe algorithmic bottlenecks ``not because it necessarily treats any individual unfairly, but because it cuts against the urgent project of scrambling existing patterns in societal inequalities''~\citep{eidelson2021patterned}. Likewise, an algorithmic monoculture can cause severe restrictions on the space of opportunity because of its potential to homogenize outcomes~\citep{kleinberg2021algorithmic, bommasani_outcome_homogenization}. Regardless of how legitimate these outcomes are considered to be, the consistency of individuals that are excluded from opportunities due to algorithmic monoculture represents a moral concern, as Creel and Hellman argue~\citep{creel_hellman_2022}. In the next section, we extend the idea that alleviating severity should be an urgent priority to systems of algorithmic decision-making and show that alleviating severity requires \textit{algorithmic pluralism}. 

\subsection{Pluralism in Algorithmic Decision-Making}

Algorithmic pluralism describes a state of affairs in which the algorithms used for decision-making are not so pervasive and/or strict as to constitute a severe bottleneck on opportunity. In defining algorithmic pluralism, we extend the goals of opportunity pluralism to systems of algorithmic decision-making. However, achieving algorithmic pluralism requires resolving an important question not fully addressed in Fishkin's text: what exactly must be plural about algorithmic decision-making? Elaborating on the principles of opportunity pluralism described in \autoref{subsec:opportunitypluralism}, pluralism in algorithmic systems could entail: 

\begin{enumerate}

\item pluralism of algorithmic components (e.g. model classes, training objectives, etc.);
\item pluralism of features or evaluation criteria used to determine who receives each opportunity~\citep[p16]{fishkin2014bottlenecks};
\item pluralism of algorithmic decision-making processes;
\item pluralism of algorithmic decision-makers~\citep[p16]{fishkin2014bottlenecks};
\item pluralism of opportunity: algorithms allow individuals with diverse characteristics (skills, identities, life paths, etc.) to have a variety of chances to gain access to valued opportunities and goods.

\end{enumerate}
In what follows, we will argue that (5) is a necessary condition for algorithmic pluralism and that while all five forms of pluralism can be valuable, (1-4) are valuable insofar as they bring about (5).

\setlength{\parindent}{0cm}
\paragraph{\textbf{These candidate types of pluralism may not be linked in systems of algorithmic decision-making.}}\quad \\
The most natural way to bring about a pluralistic opportunity structure in our society would be to encourage the proliferation of diverse institutions and decision-makers, each of which grants access to a certain kind of opportunity.  If the diverse institutions and decision-makers each make their decisions on different bases -- perhaps one evaluates numerical ability and teamwork, while another evaluates geometric and spatial reasoning and manual dexterity -- then the opportunity structure permits many different people to reach to many different valuable life paths. Therefore, all five kinds of pluralism initially appear to be linked: access to many different choice-worthy life paths\footnote{Whether meaningfully different life paths exist affects whether different people can flourish in the opportunity structure. Furthermore, the structure of opportunity can encourage people to reflect on and develop their own values and aspirations, or conversely can suppress such exploration by ensuring that only a limited number of life paths are rewarded with the basic conditions of human flourishing~\citep[p17]{fishkin2014bottlenecks}, as in the warrior society. These two forms of pluralism are important in Fishkin's account. However, it is not clear that algorithmic decision-making plays any special role in bringing about these features of the opportunity structure.  The structure of opportunity at the society level, including which social roles exist and are rewarded or dis-valued, is rarely determined by the choice of algorithms or the choice between using algorithmic and human decision-making.  Therefore we omit pluralism about societal values and goals from the types of pluralism that we consider for algorithmic decision-making.}
for (5) many different kinds of people are granted by (4) diverse institutions and (3) decision-making processes that evaluate those people on the basis of (2) many different criteria and (1) model components. For example, (5) aspiring warriors, musicians, and accountants, each with different skills and characteristics, might find opportunities at (4) the military, the orchestra, and the hedge-fund on the basis of their (2) fighting prowess, musical ability, and numerical precision, respectively. It is therefore not surprising that Fishkin consistently links these different types of pluralism in his framework.
\setlength{\parindent}{.4cm}

However, in algorithmic decision-making, these types of pluralism need not be linked. And if they are uncoupled, it is not clear what algorithmic pluralism requires. To make this vivid, imagine a society in which the pluralism(s) are not coupled: there is only one algorithm through which all must access opportunity, but since the society has vast abundance and distributes it almost equally, the top 99\% of people according to the algorithmic scores are rewarded with good material life outcomes and the ability to choose a meaningful job or life pursuit. Moreover, the algorithm is not static: it changes each year such that a different 99\% are chosen (and therefore a different 1\% are excluded). There is only one decision-maker and the algorithm applies a single evaluation criterion to everyone, so the society lacks pluralism of types (1-4). But the society supports pluralism of type (5): many different kinds of people have multiple chances to gain opportunities they value.

To put this question another way, is it enough to ensure that a diverse group of people receive opportunities they value? Or must they also be selected for those opportunities for a plurality of reasons and/or by a plurality of decision-makers? This question matters for algorithmic decision-making. Work in algorithmic fairness has investigated whether algorithmic biases create a patterned inequality~\citep{eidelson2021patterned} by acting as ``moderately strict but pervasive'' bottlenecks~\citep[p164]{fishkin2014bottlenecks} (pluralism 2) that disproportionately exclude members of marginalized subgroups (pluralism 5), but this work does not typically evaluate similarity between decision makers (pluralism 4). And work on algorithmic monoculture and outcome homogenization focuses on the extent to which similarity between decision-makers (pluralism 4) results in the same individuals being rewarded or excluded (pluralism 5) \citep{kleinberg2021algorithmic, bommasani_outcome_homogenization, creel_hellman_2022, toups2023ecosystem}, but it does not typically investigate whether the decisions are being made on the basis of the same criteria ``under the hood'' (pluralism 2).  Both literatures find pluralism (5) important, but they differ in their focuses on pluralisms (2) and (4).

\setlength{\parindent}{0cm}
\paragraph{\textbf{What types of pluralism should algorithmic pluralism require?}}\quad \\
The choice of pluralism(s) makes a significant difference to the requirements of algorithmic pluralism. If algorithmic pluralism requires that different algorithms decide based on different criteria (pluralism 2), then determining whether we are in a state of algorithmic pluralism might require access to the criteria ``on the basis of which'' each model made its decision as well as the ability to explain the relevance of each criterion. Achieving transparent access to these criteria in complex machine learning models has been notoriously challenging~\citep{Mittelstadt2019, creel2020, Gunther2021}. It would be necessary to not only achieve access to individual decision criteria, but also to have a systematic way to compare the criteria and to measure the diversity of decision criteria across the opportunity structure. 
\setlength{\parindent}{.4cm}

If algorithmic pluralism instead means ensuring that many different decision-makers control access to opportunities (pluralism 4), then we would be most interested in evaluating algorithmic monoculture and the extent to which decision-makers within the ecosystem resemble one another \citep{kleinberg2021algorithmic}. If we instead required only that the opportunity structure contains multiple different decision-making organizations, each organization could rely on the same decision-making algorithm, rendering their decisions identical to one another. In order for pluralism (4) to be meaningful, the decision-makers must be different at the level of outcomes, not merely different in organization name. 

Pluralism of opportunity (5) requires that the opportunity structure be devoid of severe bottlenecks so that a broad set of individuals can access opportunity. The winners are not the only ones who can keep winning; those unfortunate in one contest have other chances to flourish in the future. As Creel and Hellman argue, hiring algorithms can achieve this kind of pluralism by allowing different kinds of job candidates to succeed, neither consistently making mistakes on the same individuals nor establishing only one criterion for their success~\citep{creel_hellman_2022}. Due to the consistency of algorithmic decision-making, this type of algorithmic pluralism is especially important in ensuring that many individuals have access to the opportunities that would allow them to flourish. 

\subsection{Algorithmic Pluralism Requires Pluralism of Opportunity}

We argue that only pluralism of opportunity (5) is \textit{necessary} for algorithmic pluralism. If we ultimately hope to alleviate severe bottlenecks and promote opportunity pluralism, a range of people must actually have access to and receive different forms of opportunity that they value. In other words, algorithmic systems must allow a pluralism of opportunity (5). Pluralisms (1-4) are \textit{instrumentally} valuable, which is to say they are valuable only insofar as they achieve the goal of (5). Having diverse model components and evaluation criteria across different decision-makers and processes is not valuable on its own -- it is only valuable if it leads to diverse model outcomes, each of which opens doors to opportunity for different people. If the diverse models all deliver opportunity to the same small group of people, the diversity of their individual components is not materially meaningful. In short, algorithmic pluralism requires algorithms to differ from one another in their decision-outcomes, often because they differ in the bases for their decisions, allowing decision-subjects multiple kinds of opportunity. 

Achieving pluralism of opportunity (5) in practice might require bringing about one or more of pluralisms (1-4): algorithms used by different decision-makers (4) that differ in their components (1), in their evaluation criteria (2), or in the broader decision-making processes surrounding them that lead to outcomes (3). As we will discuss in Section~\ref{sec:implications}, regulators and designers should often seek to achieve pluralism by promoting (1-4). However, it is also possible for a single algorithm to bring about (5) if (for example) it incorporates randomness in its decision-making to broaden the diversity of people who are given access to valued outcomes~\citep{creel_hellman_2022, jain2024scarce}. What ultimately matters is that the resulting opportunity structure is meaningfully pluralistic in offering a diverse set of individuals with access to opportunity. In the next section, we use a case study of algorithmic hiring to further illustrate the idea of algorithmic pluralism.

\section{Case Study: Algorithmic Hiring}\label{sec:caseStudies}
We elaborate on the importance of algorithmic pluralism with a case study of algorithmic hiring. Algorithms constitute severe bottlenecks across all stages of contemporary hiring. Resume-screening and skill-assessment algorithms often reject the majority of applicants, making them strict bottlenecks on the path to receiving an interview. Moreover, most Fortune 500 companies rely on hiring tools from third-party vendors such as HireVue and Pymetrics~\citep{raghavan2020hiring}, creating the risk of the same algorithms being pervasive bottlenecks for multiple jobs, strictly excluding the same candidates. The recent advances in generative AI may increase the adoption of similar automated tools across the labor market~\citep{Kaashoek2024Impact}.

Hiring algorithms have the potential to reinforce existing patterns of inequality in the employment sector. Decades of audit studies show that employers tend to discriminate against racial and gender minorities, with little improvement over the past 25 years~\citep{raghavan2020hiring}. Unsurprisingly, Amazon's original resume-screening algorithm downgraded applicants from women's colleges, in part because it was trained on its existing majority-male employee base~\citep{amazonBias}. LinkedIn's job recommendation algorithms also ended up referring more men than women for open roles because it tried to maximize applicants and men were more likely to apply even if they didn't meet the qualifications~\citep{linkedinBias}. In addition to replicating the biases long demonstrated in human hiring, algorithmic hiring has the potential to establish a uniquely severe bottleneck to employment. If many employers rely on the same large third-party provider of candidate screening software, the same individual candidates or subgroups could be consistently and erroneously rejected~\citep{bommasani_outcome_homogenization}. This targeted exclusion could be both severe and illegitimate due to the arbitrariness of the factors on which candidates are rejected~\citep{creel_hellman_2022}.  

Disparate impact regulations in the United States aim to loosen bottlenecks for legally protected groups, but existing regulations fall short of alleviating the bottleneck imposed by hiring algorithms. The U.S. Equal Employment Opportunity Commission (EEOC) has established the 4/5 rule as precedence for when a disparate impact case can be brought against an employer~\citep{eeocHiring}. The rule takes a formal view on equal opportunity and states that employers may be at legal risk of discrimination if the selection rate for one protected group (e.g. race, gender, disability, age, religion, etc.) is less than 4/5 of that of the group with the highest selection rate. However, employers can still maintain their selection procedures by showing a ``business necessity''~\citep{eeocHiring}. This often leads to zero-sum debates in the courts over what procedures are legitimate, where the status-quo benefits one group and the alternatives would distribute opportunities to another group. Neither option offered by the current regulations fully addresses the strictness or pervasiveness of bottlenecks created by algorithmic hiring procedures.

The EEOC could adopt several interventions in order to promote algorithmic pluralism. As third parties with the power to audit and monitor firms, regulators are in the best position to measure the extent to which a pluralism of opportunity (pluralism 5) exists across a job sector. Specifically, the EEOC should audit the homogeneity of hiring outcomes by collecting information about which candidates are rejected and the extent to which these outcomes are systemic and algorithmically driven. While the EEOC recommends that companies should analyze and audit the automated hiring tools they use~\citep{eeocAlgorithms}, individual firms cannot fully observe the decisions made by their competitors. For example, an individual firm may not find it concerning if their resume-screening algorithm rejects individuals who have employment gaps in their resume. But all companies might use algorithms that reject these individuals, even if they had to stop working for legitimate reasons like childcare or health problems, resulting in the concentration of unfair treatment. Third-party audits could help uncover groups of unemployed persons that are systemically rejected and motivate positive-sum protections for these groups, such as ways for candidates with special circumstances to flag these in their application (pluralism 3). Furthermore, if audits  uncover algorithmic monocultures, the Federal Trade Commission (FTC) should consider stepping in to prevent anti-competitive practices, especially if direct competitors in a job sector are using the same hiring algorithms (pluralism 4).   

Employers can further shift their hiring processes towards algorithmic pluralism and have several incentives to do so. If all competitors rely on the same hiring algorithm, then the overall quality of hired applicants can decrease over time because those selected by the algorithm can only choose one company for which to work~\citep{kleinberg2021algorithmic, peng2023monoculture}. Employers may also have different conceptions of applicant merit or business priorities that a standardized algorithm will fail to capture. To address these problems, companies could modify the design of hiring algorithms to prioritize certain features such as skills on a resume (pluralism 2) or add different training objectives like hiring from the local community (pluralism 1). Moreover, many algorithms will undervalue applicants from under-represented groups and fail to learn about changes in applicant hiring potential over time. This should incentivize employers to not strictly follow algorithmic rankings (pluralism 3) and balance exploitation (selecting individuals with proven track records) with exploration (selecting individuals about which the predictive algorithm is less certain in order to learn)~\citep{li2020hiring}. The inherent uncertainty in evaluating candidate merit may also motivate some use of randomization in selecting applicants~\citep{singh2021fairness, jain2024scarce}. These interventions all lead to what ultimately matters for algorithmic pluralism: allowing applicants with diverse characteristics (pluralism 5) to make it through the bottlenecks of hiring algorithms and have meaningful chances at job opportunities.

\section{Implications for Algorithmic Regulation and Design}\label{sec:implications}
Whether in hiring or in other domains, regulators and policymakers are often in the best position to identify severe bottlenecks and measure the extent to which pluralism of opportunity is compromised.
While individual decision-makers may not be able to fully observe the bottlenecks in their ecosystem, they can still create pluralistic algorithms and may even have incentives to do so. 

\subsection{Regulators and Policymakers}
Regulators and policymakers in many branches of government are explicitly tasked with equalizing the structure of opportunity. For algorithmic systems in particular, this may first require the development of mechanisms to quantify the severity of bottlenecks. One such metric could involve the extent to which deployed models make homogeneous predictions~\citep{bommasani_outcome_homogenization, toups2023ecosystem}. However, many current systems of decision-making lack transparency about what stages of the process are algorithmically-mediated and which features of individuals are the basis of negative decisions that deny them opportunities. Regulators have the unique capacity to collect this information through audits or mandated reporting while still protecting privacy and competitive advantages. These audits could also help uncover the pervasiveness of certain algorithms or models and the extent to which algorithmic monocultures are present.  

U.S. anti-discrimination and antitrust laws offer ways to promote pluralism in algorithmic systems that amount to a severe bottleneck. In some domains, algorithmic monocultures contribute to anti-competitive practices that harm consumers and warrant legal intervention. For example, recent lawsuits allege that the shared use of a rent-pricing algorithm (RealPage) enables landlords to fix apartment prices illegally~\citep{wsjRealPage}. These cases offer a precedent for how antitrust laws may be used to ensure a pluralism of algorithmic decision-makers (pluralism 4). Many federal and state anti-discrimination laws also offer ways to address monocultures in algorithmic decision-making processes (pluralism 3) so that bottlenecks are somewhat less severe. Legal protections exist for groups to whom we would not ordinarily expect the law to show solicitude, such as those with low-credit, predispositions to disease, and even prior criminal convictions~\citep{fishkin2014bottlenecks}. These protections prohibit the use of certain information in earlier stages of decision-making (e.g. the screening stage for job interviews)~\citep{ballance2020creditScore} and also create avenues for recourse or redress~\citep{venkatasubramanian2020recourse}. The latter form of protection is especially important for algorithmic decision-making systems because there is often a lack of transparency that makes it difficult to identify those whose opportunities are severely constrained.

\subsection{Designers and Decision-Makers}

Algorithmic pluralism calls for a state of affairs in which the decision-making algorithms in a sector allow many different individuals to receive chances at valued opportunities and goods (pluralism 5). In practice, individual decision-makers can make many model and system design choices that help bring about algorithmic pluralism. For example, decision-makers may want to embed their own unique values and preferences in the choice of features and evaluation criteria (pluralism 2) or stages of the process that are algorithmically-mediated (pluralism 3). Even for the same problem formulation, different choices of model components (pluralism 1) can all yield similar accuracy but varying predictions -- a phenomenon known as ``predictive multiplicity''~\citep{marx2020predictive} or ``model multiplicity''~\citep{black_multiplicity}. In market settings where there is a competition for the same candidates, these forms of pluralism can lead to better outcomes for individual decision-makers~\citep{kleinberg2021algorithmic, peng2023monoculture}. Some domains may also impose a legal duty on decision-makers to search for ``less discriminatory algorithms'', or model components that lead to outcomes with fewer disparate impacts across demographic groups~\citep{black2023less}. 

If different decision-makers adopt pluralism in their design choices, it increases the likelihood of satisfying algorithmic pluralism in decision outcomes. However, in some settings different design choices can still result in correlated outcomes~\citep{bommasani_outcome_homogenization, toups2023ecosystem}, while in others there may be only one decision-maker or algorithm in use. For these settings, we affirm proposals to intentionally introduce randomness into the decision-making process~\citep{creel_hellman_2022, jain2024scarce}. These settings often involve an inherent uncertainty in predictions. Predictive uncertainty justifies introducing principled randomization within these bounds of uncertainty that does not involve a loss in utility for the decision-maker~\citep{singh2021fairness, jain2024scarce}. We encourage future work to develop technical solutions that can help disrupt the severity of decisions and promote algorithmic pluralism.


\section{Concluding Remarks}\label{sec:conclusion}
Fishkin's idea of opportunity pluralism makes the case that societies ought to move their structures of opportunity towards a more pluralistic model, where there are many gatekeepers and paths towards opportunities~\citep{fishkin2014bottlenecks}. As our economy increasingly relies on artificial intelligence, we emphasize the importance of extending this idea to systems of data-driven decision-making through algorithmic pluralism. Towards the end of his book, Fishkin emphasizes that opportunity pluralism has vast implications for various institutions and stakeholders. He aspires to encourage gatekeepers who wish to help build a more pluralistic opportunity structure to re-examine and ameliorate the bottlenecks that result from how they make decisions. We similarly implore the data scientists responsible for designing models, gatekeepers who ultimately make decisions, and policymakers regulating algorithmic systems to determine where they have the leverage to ameliorate bottlenecks and help promote a pluralism of opportunity for all.

\clearpage
\bibliographystyle{ACM-Reference-Format}
\bibliography{refs}


\begin{thebibliography}{59}


\ifx \showCODEN    \undefined \def \showCODEN     #1{\unskip}     \fi
\ifx \showDOI      \undefined \def \showDOI       #1{#1}\fi
\ifx \showISBNx    \undefined \def \showISBNx     #1{\unskip}     \fi
\ifx \showISBNxiii \undefined \def \showISBNxiii  #1{\unskip}     \fi
\ifx \showISSN     \undefined \def \showISSN      #1{\unskip}     \fi
\ifx \showLCCN     \undefined \def \showLCCN      #1{\unskip}     \fi
\ifx \shownote     \undefined \def \shownote      #1{#1}          \fi
\ifx \showarticletitle \undefined \def \showarticletitle #1{#1}   \fi
\ifx \showURL      \undefined \def \showURL       {\relax}        \fi
\providecommand\bibfield[2]{#2}
\providecommand\bibinfo[2]{#2}
\providecommand\natexlab[1]{#1}
\providecommand\showeprint[2][]{arXiv:#2}

\bibitem[Ajunwa(2019)]%
        {Ajunwa2021}
\bibfield{author}{\bibinfo{person}{Ifeoma Ajunwa}.}
  \bibinfo{year}{2019}\natexlab{}.
\newblock \showarticletitle{Automated employment discrimination}.
\newblock \bibinfo{journal}{\emph{Harvard Journal of Law and Technology}}
  \bibinfo{volume}{34} (\bibinfo{year}{2019}), \bibinfo{pages}{622 -- 699}.
\newblock


\bibitem[Anderson(1999)]%
        {Anderson1999}
\bibfield{author}{\bibinfo{person}{Elizabeth~S. Anderson}.}
  \bibinfo{year}{1999}\natexlab{}.
\newblock \showarticletitle{What Is the Point of Equality?}
\newblock \bibinfo{journal}{\emph{Ethics}} \bibinfo{volume}{109},
  \bibinfo{number}{2} (\bibinfo{date}{Jan.} \bibinfo{year}{1999}),
  \bibinfo{pages}{287--337}.
\newblock
\urldef\tempurl%
\url{https://doi.org/10.1086/233897}
\showDOI{\tempurl}


\bibitem[Arif~Khan et~al\mbox{.}(2022)]%
        {eaamo2022equal}
\bibfield{author}{\bibinfo{person}{Falaah Arif~Khan}, \bibinfo{person}{Eleni
  Manis}, {and} \bibinfo{person}{Julia Stoyanovich}.}
  \bibinfo{year}{2022}\natexlab{}.
\newblock \showarticletitle{Towards Substantive Conceptions of Algorithmic
  Fairness: Normative Guidance from Equal Opportunity Doctrines}. In
  \bibinfo{booktitle}{\emph{Equity and Access in Algorithms, Mechanisms, and
  Optimization}} (Arlington, VA, USA) \emph{(\bibinfo{series}{EAAMO '22})}.
  \bibinfo{publisher}{Association for Computing Machinery},
  \bibinfo{address}{New York, NY, USA}, Article \bibinfo{articleno}{18},
  \bibinfo{numpages}{10}~pages.
\newblock
\showISBNx{9781450394772}
\urldef\tempurl%
\url{https://doi.org/10.1145/3551624.3555303}
\showDOI{\tempurl}


\bibitem[Ballance et~al\mbox{.}(2020)]%
        {ballance2020creditScore}
\bibfield{author}{\bibinfo{person}{Joshua Ballance}, \bibinfo{person}{Robert
  Clifford}, {and} \bibinfo{person}{Daniel Shoag}.}
  \bibinfo{year}{2020}\natexlab{}.
\newblock \showarticletitle{“No more credit score”: Employer credit check
  bans and signal substitution}.
\newblock \bibinfo{journal}{\emph{Labour Economics}}  \bibinfo{volume}{63}
  (\bibinfo{year}{2020}), \bibinfo{pages}{101769}.
\newblock


\bibitem[Black et~al\mbox{.}(2024)]%
        {black2023less}
\bibfield{author}{\bibinfo{person}{Emily Black}, \bibinfo{person}{John~Logan
  Koepke}, \bibinfo{person}{Pauline Kim}, \bibinfo{person}{Solon Barocas},
  {and} \bibinfo{person}{Mingwei Hsu}.} \bibinfo{year}{2024}\natexlab{}.
\newblock \bibinfo{title}{Less Discriminatory Algorithms}.
\newblock
\newblock


\bibitem[Black et~al\mbox{.}(2022)]%
        {black_multiplicity}
\bibfield{author}{\bibinfo{person}{Emily Black}, \bibinfo{person}{Manish
  Raghavan}, {and} \bibinfo{person}{Solon Barocas}.}
  \bibinfo{year}{2022}\natexlab{}.
\newblock \showarticletitle{Model Multiplicity: Opportunities, Concerns, and
  Solutions}. In \bibinfo{booktitle}{\emph{2022 ACM Conference on Fairness,
  Accountability, and Transparency}} (Seoul, Republic of Korea)
  \emph{(\bibinfo{series}{FAccT '22})}. \bibinfo{publisher}{Association for
  Computing Machinery}, \bibinfo{address}{New York, NY, USA},
  \bibinfo{pages}{850–863}.
\newblock
\showISBNx{9781450393522}
\urldef\tempurl%
\url{https://doi.org/10.1145/3531146.3533149}
\showDOI{\tempurl}


\bibitem[Bommasani et~al\mbox{.}(2022)]%
        {bommasani_outcome_homogenization}
\bibfield{author}{\bibinfo{person}{Rishi Bommasani},
  \bibinfo{person}{Kathleen~A Creel}, \bibinfo{person}{Ananya Kumar},
  \bibinfo{person}{Dan Jurafsky}, {and} \bibinfo{person}{Percy~S Liang}.}
  \bibinfo{year}{2022}\natexlab{}.
\newblock \showarticletitle{Picking on the same person: Does algorithmic
  monoculture lead to outcome homogenization?}
\newblock \bibinfo{journal}{\emph{Advances in Neural Information Processing
  Systems}}  \bibinfo{volume}{35} (\bibinfo{year}{2022}),
  \bibinfo{pages}{3663--3678}.
\newblock


\bibitem[Bommasani et~al\mbox{.}(2021)]%
        {bommasani2021opportunities}
\bibfield{author}{\bibinfo{person}{Rishi Bommasani}, \bibinfo{person}{Drew~A
  Hudson}, \bibinfo{person}{Ehsan Adeli}, \bibinfo{person}{Russ Altman},
  \bibinfo{person}{Simran Arora}, \bibinfo{person}{Sydney von Arx},
  \bibinfo{person}{Michael~S Bernstein}, \bibinfo{person}{Jeannette Bohg},
  \bibinfo{person}{Antoine Bosselut}, \bibinfo{person}{Emma Brunskill},
  {et~al\mbox{.}}} \bibinfo{year}{2021}\natexlab{}.
\newblock \bibinfo{title}{On the opportunities and risks of foundation models}.
\newblock
\newblock


\bibitem[Calders et~al\mbox{.}(2009)]%
        {calders2009building}
\bibfield{author}{\bibinfo{person}{Toon Calders}, \bibinfo{person}{Faisal
  Kamiran}, {and} \bibinfo{person}{Mykola Pechenizkiy}.}
  \bibinfo{year}{2009}\natexlab{}.
\newblock \showarticletitle{Building classifiers with independency
  constraints}. In \bibinfo{booktitle}{\emph{2009 IEEE international conference
  on data mining workshops}}. \bibinfo{publisher}{IEEE},
  \bibinfo{address}{IEEE}, \bibinfo{pages}{13--18}.
\newblock


\bibitem[Commission(1979)]%
        {eeocHiring}
\bibfield{author}{\bibinfo{person}{US~Equal Employment~Opportunity
  Commission}.} \bibinfo{year}{1979}\natexlab{}.
\newblock \bibinfo{title}{Questions and Answers to Clarify and Provide a Common
  Interpretation of the Uniform Guidelines on Employee Selection Procedures}.
\newblock
\newblock


\bibitem[Commission(2023)]%
        {eeocAlgorithms}
\bibfield{author}{\bibinfo{person}{US~Equal Employment~Opportunity
  Commission}.} \bibinfo{year}{2023}\natexlab{}.
\newblock \bibinfo{title}{Select Issues: Assessing Adverse Impact in Software,
  Algorithms, and Artificial Intelligence Used in Employment Selection
  Procedures Under Title VII of the Civil Rights Act of 1964}.
\newblock
\newblock


\bibitem[Corbett-Davies et~al\mbox{.}(2016)]%
        {corbett2016COMPAS}
\bibfield{author}{\bibinfo{person}{Sam Corbett-Davies}, \bibinfo{person}{Emma
  Pierson}, \bibinfo{person}{Avi Feller}, {and} \bibinfo{person}{Sharad Goel}.}
  \bibinfo{year}{2016}\natexlab{}.
\newblock \bibinfo{title}{A computer program used for bail and sentencing
  decisions was labeled biased against blacks. It’s actually not that clear}.
\newblock
\newblock


\bibitem[Costanza-Chock(2018)]%
        {costanza2018design}
\bibfield{author}{\bibinfo{person}{Sasha Costanza-Chock}.}
  \bibinfo{year}{2018}\natexlab{}.
\newblock \showarticletitle{Design justice, AI, and escape from the matrix of
  domination}.
\newblock \bibinfo{journal}{\emph{Journal of Design and Science}}
  \bibinfo{volume}{3}, \bibinfo{number}{5} (\bibinfo{year}{2018}),
  \bibinfo{pages}{1--14}.
\newblock


\bibitem[Creel and Hellman(2022)]%
        {creel_hellman_2022}
\bibfield{author}{\bibinfo{person}{Kathleen Creel} {and}
  \bibinfo{person}{Deborah Hellman}.} \bibinfo{year}{2022}\natexlab{}.
\newblock \showarticletitle{The Algorithmic Leviathan: Arbitrariness, Fairness,
  and Opportunity in Algorithmic Decision-Making Systems}.
\newblock \bibinfo{journal}{\emph{Canadian Journal of Philosophy}}
  \bibinfo{volume}{52}, \bibinfo{number}{1} (\bibinfo{year}{2022}),
  \bibinfo{pages}{26–43}.
\newblock
\urldef\tempurl%
\url{https://doi.org/10.1017/can.2022.3}
\showDOI{\tempurl}


\bibitem[Creel(2020)]%
        {creel2020}
\bibfield{author}{\bibinfo{person}{Kathleen~A. Creel}.}
  \bibinfo{year}{2020}\natexlab{}.
\newblock \showarticletitle{Transparency in Complex Computational Systems}.
\newblock \bibinfo{journal}{\emph{Philosophy of Science}} \bibinfo{volume}{87},
  \bibinfo{number}{4} (\bibinfo{date}{Oct.} \bibinfo{year}{2020}),
  \bibinfo{pages}{568--589}.
\newblock
\urldef\tempurl%
\url{https://doi.org/10.1086/709729}
\showDOI{\tempurl}


\bibitem[Davis et~al\mbox{.}(2021)]%
        {davis2021reparation}
\bibfield{author}{\bibinfo{person}{Jenny~L Davis}, \bibinfo{person}{Apryl
  Williams}, {and} \bibinfo{person}{Michael~W Yang}.}
  \bibinfo{year}{2021}\natexlab{}.
\newblock \showarticletitle{Algorithmic reparation}.
\newblock \bibinfo{journal}{\emph{Big Data \& Society}} \bibinfo{volume}{8},
  \bibinfo{number}{2} (\bibinfo{year}{2021}),
  \bibinfo{pages}{20539517211044808}.
\newblock


\bibitem[Dewey(2018)]%
        {dewey1938logic}
\bibfield{author}{\bibinfo{person}{John Dewey}.}
  \bibinfo{year}{2018}\natexlab{}.
\newblock \bibinfo{booktitle}{\emph{Logic-The theory of inquiry}}.
\newblock \bibinfo{publisher}{Read Books Ltd}, \bibinfo{address}{Redditch ,
  Great Britain}.
\newblock


\bibitem[Eidelson(2021)]%
        {eidelson2021patterned}
\bibfield{author}{\bibinfo{person}{Benjamin Eidelson}.}
  \bibinfo{year}{2021}\natexlab{}.
\newblock \showarticletitle{Patterned Inequality, Compounding Injustice, and
  Algorithmic Prediction}.
\newblock \bibinfo{journal}{\emph{American Journal of Law and Equality}}
  \bibinfo{volume}{1} (\bibinfo{year}{2021}), \bibinfo{pages}{252--276}.
\newblock


\bibitem[Fishkin(2014)]%
        {fishkin2014bottlenecks}
\bibfield{author}{\bibinfo{person}{Joseph Fishkin}.}
  \bibinfo{year}{2014}\natexlab{}.
\newblock \bibinfo{booktitle}{\emph{Bottlenecks: A new theory of equal
  opportunity}}.
\newblock \bibinfo{publisher}{Oxford University Press, USA},
  \bibinfo{address}{New York, NY}.
\newblock


\bibitem[Fishkin and Forbath(2014)]%
        {fishkin2014oligarchy}
\bibfield{author}{\bibinfo{person}{Joseph Fishkin} {and}
  \bibinfo{person}{William~E Forbath}.} \bibinfo{year}{2014}\natexlab{}.
\newblock \showarticletitle{The anti-oligarchy constitution}.
\newblock \bibinfo{journal}{\emph{BUL Rev.}}  \bibinfo{volume}{94}
  (\bibinfo{year}{2014}), \bibinfo{pages}{669}.
\newblock


\bibitem[Friedler et~al\mbox{.}(2021)]%
        {friedler2021possibility}
\bibfield{author}{\bibinfo{person}{Sorelle~A Friedler}, \bibinfo{person}{Carlos
  Scheidegger}, {and} \bibinfo{person}{Suresh Venkatasubramanian}.}
  \bibinfo{year}{2021}\natexlab{}.
\newblock \showarticletitle{The (im) possibility of fairness: Different value
  systems require different mechanisms for fair decision making}.
\newblock \bibinfo{journal}{\emph{Commun. ACM}} \bibinfo{volume}{64},
  \bibinfo{number}{4} (\bibinfo{year}{2021}), \bibinfo{pages}{136--143}.
\newblock


\bibitem[Green(2022)]%
        {Green2022}
\bibfield{author}{\bibinfo{person}{Ben Green}.}
  \bibinfo{year}{2022}\natexlab{}.
\newblock \showarticletitle{Escaping the impossibility of fairness: From formal
  to substantive algorithmic fairness}.
\newblock \bibinfo{journal}{\emph{Philosophy \& Technology}}
  \bibinfo{volume}{35}, \bibinfo{number}{4} (\bibinfo{year}{2022}),
  \bibinfo{pages}{90}.
\newblock


\bibitem[Green and Viljoen(2020)]%
        {green_realism}
\bibfield{author}{\bibinfo{person}{Ben Green} {and} \bibinfo{person}{Salom\'{e}
  Viljoen}.} \bibinfo{year}{2020}\natexlab{}.
\newblock \showarticletitle{Algorithmic Realism: Expanding the Boundaries of
  Algorithmic Thought}. In \bibinfo{booktitle}{\emph{Proceedings of the 2020
  Conference on Fairness, Accountability, and Transparency}} (Barcelona, Spain)
  \emph{(\bibinfo{series}{FAT* '20})}. \bibinfo{publisher}{Association for
  Computing Machinery}, \bibinfo{address}{New York, NY, USA},
  \bibinfo{pages}{19–31}.
\newblock
\showISBNx{9781450369367}
\urldef\tempurl%
\url{https://doi.org/10.1145/3351095.3372840}
\showDOI{\tempurl}


\bibitem[G\"{u}nther and Kasirzadeh(2021)]%
        {Gunther2021}
\bibfield{author}{\bibinfo{person}{Mario G\"{u}nther} {and}
  \bibinfo{person}{Atoosa Kasirzadeh}.} \bibinfo{year}{2021}\natexlab{}.
\newblock \showarticletitle{Algorithmic and human decision making: for a double
  standard of transparency}.
\newblock \bibinfo{journal}{\emph{{AI} \& {SOCIETY}}} \bibinfo{volume}{37},
  \bibinfo{number}{1} (\bibinfo{date}{April} \bibinfo{year}{2021}),
  \bibinfo{pages}{375--381}.
\newblock
\urldef\tempurl%
\url{https://doi.org/10.1007/s00146-021-01200-5}
\showDOI{\tempurl}


\bibitem[Hardt and Kim(2022)]%
        {hardt2022backward}
\bibfield{author}{\bibinfo{person}{Moritz Hardt} {and}
  \bibinfo{person}{Michael~P Kim}.} \bibinfo{year}{2022}\natexlab{}.
\newblock \bibinfo{title}{Backward baselines: Is your model predicting the
  past?}
\newblock
\newblock


\bibitem[Hardt et~al\mbox{.}(2016)]%
        {hardt2016equality}
\bibfield{author}{\bibinfo{person}{Moritz Hardt}, \bibinfo{person}{Eric Price},
  {and} \bibinfo{person}{Nati Srebro}.} \bibinfo{year}{2016}\natexlab{}.
\newblock \bibinfo{title}{Equality of opportunity in supervised learning}.
\newblock
\newblock


\bibitem[H{\'e}bert-Johnson et~al\mbox{.}(2018)]%
        {hebert-johnsonMulticalibration}
\bibfield{author}{\bibinfo{person}{Ursula H{\'e}bert-Johnson},
  \bibinfo{person}{Michael Kim}, \bibinfo{person}{Omer Reingold}, {and}
  \bibinfo{person}{Guy Rothblum}.} \bibinfo{year}{2018}\natexlab{}.
\newblock \showarticletitle{Multicalibration: Calibration for the
  (computationally-identifiable) masses}. In
  \bibinfo{booktitle}{\emph{International Conference on Machine Learning}}.
  \bibinfo{publisher}{PMLR}, \bibinfo{address}{Vienna, Austria},
  \bibinfo{pages}{1939--1948}.
\newblock


\bibitem[Hellman(2018)]%
        {hellman2018compounding}
\bibfield{author}{\bibinfo{person}{Deborah Hellman}.}
  \bibinfo{year}{2018}\natexlab{}.
\newblock \bibinfo{title}{Indirect discrimination and the duty to avoid
  compounding injustice}.
\newblock , \bibinfo{numpages}{2017--53}~pages.
\newblock


\bibitem[Hoffmann(2019)]%
        {hoffmann2019fairness}
\bibfield{author}{\bibinfo{person}{Anna~Lauren Hoffmann}.}
  \bibinfo{year}{2019}\natexlab{}.
\newblock \showarticletitle{Where fairness fails: data, algorithms, and the
  limits of antidiscrimination discourse}.
\newblock \bibinfo{journal}{\emph{Information, Communication \& Society}}
  \bibinfo{volume}{22}, \bibinfo{number}{7} (\bibinfo{year}{2019}),
  \bibinfo{pages}{900--915}.
\newblock


\bibitem[Holm(2023)]%
        {Holm2023}
\bibfield{author}{\bibinfo{person}{Sune Holm}.}
  \bibinfo{year}{2023}\natexlab{}.
\newblock \showarticletitle{Algorithmic legitimacy in clinical
  decision-making}.
\newblock \bibinfo{journal}{\emph{Ethics and Information Technology}}
  \bibinfo{volume}{25}, \bibinfo{number}{3} (\bibinfo{year}{2023}),
  \bibinfo{pages}{35}.
\newblock


\bibitem[Jain et~al\mbox{.}(2024)]%
        {jain2024scarce}
\bibfield{author}{\bibinfo{person}{Shomik Jain}, \bibinfo{person}{Kathleen
  Creel}, {and} \bibinfo{person}{Ashia Wilson}.}
  \bibinfo{year}{2024}\natexlab{}.
\newblock \showarticletitle{Scarce Resource Allocations That Rely On Machine
  Learning Should Be Randomized}. In \bibinfo{booktitle}{\emph{International
  Conference on Machine Learning}}. \bibinfo{publisher}{PMLR},
  \bibinfo{address}{Vienna, Austria}.
\newblock


\bibitem[Kaashoek et~al\mbox{.}(2024)]%
        {Kaashoek2024Impact}
\bibfield{author}{\bibinfo{person}{Justin Kaashoek}, \bibinfo{person}{Manish
  Raghavan}, {and} \bibinfo{person}{John~J. Horton}.}
  \bibinfo{year}{2024}\natexlab{}.
\newblock \bibinfo{title}{The {Impact} of {Generative} {AI} on {Labor} {Market}
  {Matching}}.
\newblock
\newblock
\newblock
\shownote{https://mit-genai.pubpub.org/pub/4t8pqt06}.


\bibitem[Kasirzadeh(2022)]%
        {kasirzadeh2022feminist}
\bibfield{author}{\bibinfo{person}{Atoosa Kasirzadeh}.}
  \bibinfo{year}{2022}\natexlab{}.
\newblock \showarticletitle{Algorithmic Fairness and Structural Injustice:
  Insights from Feminist Political Philosophy}. In
  \bibinfo{booktitle}{\emph{Proceedings of the 2022 AAAI/ACM Conference on AI,
  Ethics, and Society}}. \bibinfo{publisher}{AAAI}, \bibinfo{address}{Oxford,
  England}, \bibinfo{pages}{349--356}.
\newblock


\bibitem[Kasy and Abebe(2021)]%
        {kasy2021fairness}
\bibfield{author}{\bibinfo{person}{Maximilian Kasy} {and}
  \bibinfo{person}{Rediet Abebe}.} \bibinfo{year}{2021}\natexlab{}.
\newblock \showarticletitle{Fairness, equality, and power in algorithmic
  decision-making}. In \bibinfo{booktitle}{\emph{Proceedings of the 2021 ACM
  Conference on Fairness, Accountability, and Transparency}}.
  \bibinfo{publisher}{ACM}, \bibinfo{address}{Virtual},
  \bibinfo{pages}{576--586}.
\newblock


\bibitem[Kleinberg et~al\mbox{.}(2016)]%
        {kleinberg2016inherent}
\bibfield{author}{\bibinfo{person}{Jon Kleinberg}, \bibinfo{person}{Sendhil
  Mullainathan}, {and} \bibinfo{person}{Manish Raghavan}.}
  \bibinfo{year}{2016}\natexlab{}.
\newblock \bibinfo{title}{Inherent trade-offs in the fair determination of risk
  scores}.
\newblock
\newblock


\bibitem[Kleinberg and Raghavan(2021)]%
        {kleinberg2021algorithmic}
\bibfield{author}{\bibinfo{person}{Jon Kleinberg} {and} \bibinfo{person}{Manish
  Raghavan}.} \bibinfo{year}{2021}\natexlab{}.
\newblock \showarticletitle{Algorithmic monoculture and social welfare}.
\newblock \bibinfo{journal}{\emph{Proceedings of the National Academy of
  Sciences}} \bibinfo{volume}{118}, \bibinfo{number}{22}
  (\bibinfo{year}{2021}), \bibinfo{pages}{e2018340118}.
\newblock


\bibitem[Li et~al\mbox{.}(2020)]%
        {li2020hiring}
\bibfield{author}{\bibinfo{person}{Danielle Li}, \bibinfo{person}{Lindsey~R
  Raymond}, {and} \bibinfo{person}{Peter Bergman}.}
  \bibinfo{year}{2020}\natexlab{}.
\newblock \bibinfo{booktitle}{\emph{Hiring as exploration}}.
\newblock \bibinfo{type}{{T}echnical {R}eport}. \bibinfo{institution}{National
  Bureau of Economic Research}.
\newblock


\bibitem[Marx et~al\mbox{.}(2020)]%
        {marx2020predictive}
\bibfield{author}{\bibinfo{person}{Charles Marx}, \bibinfo{person}{Flavio
  Calmon}, {and} \bibinfo{person}{Berk Ustun}.}
  \bibinfo{year}{2020}\natexlab{}.
\newblock \showarticletitle{Predictive multiplicity in classification}. In
  \bibinfo{booktitle}{\emph{International Conference on Machine Learning}}.
  PMLR, \bibinfo{publisher}{ACM}, \bibinfo{address}{Virtual},
  \bibinfo{pages}{6765--6774}.
\newblock


\bibitem[Minow(1990)]%
        {minow1990making}
\bibfield{author}{\bibinfo{person}{Martha Minow}.}
  \bibinfo{year}{1990}\natexlab{}.
\newblock \bibinfo{booktitle}{\emph{Making all the difference: Inclusion,
  exclusion, and American law}}.
\newblock \bibinfo{publisher}{Cornell University Press},
  \bibinfo{address}{Ithaca, NY}.
\newblock


\bibitem[Mittelstadt et~al\mbox{.}(2019)]%
        {Mittelstadt2019}
\bibfield{author}{\bibinfo{person}{Brent Mittelstadt}, \bibinfo{person}{Chris
  Russell}, {and} \bibinfo{person}{Sandra Wachter}.}
  \bibinfo{year}{2019}\natexlab{}.
\newblock \showarticletitle{Explaining explanations in AI}. In
  \bibinfo{booktitle}{\emph{Proceedings of the Conference on Fairness,
  Accountability, and Transparency}}. \bibinfo{publisher}{ACM},
  \bibinfo{address}{Atlanta, GA}, \bibinfo{pages}{279--288}.
\newblock


\bibitem[Nawrat(2023)]%
        {hirevueArticle}
\bibfield{author}{\bibinfo{person}{Allie Nawrat}.}
  \bibinfo{year}{2023}\natexlab{}.
\newblock \bibinfo{title}{Inside HireVue's acquisition of Modern Hire}.
\newblock
  \bibinfo{howpublished}{\url{https://www.unleash.ai/hr-technology/inside-hirevues-acquisition-of-modern-hire/}}.
\newblock


\bibitem[Obermeyer et~al\mbox{.}(2019)]%
        {obermeyer2019proxy}
\bibfield{author}{\bibinfo{person}{Ziad Obermeyer}, \bibinfo{person}{Brian
  Powers}, \bibinfo{person}{Christine Vogeli}, {and} \bibinfo{person}{Sendhil
  Mullainathan}.} \bibinfo{year}{2019}\natexlab{}.
\newblock \showarticletitle{Dissecting racial bias in an algorithm used to
  manage the health of populations}.
\newblock \bibinfo{journal}{\emph{Science}} \bibinfo{volume}{366},
  \bibinfo{number}{6464} (\bibinfo{year}{2019}), \bibinfo{pages}{447--453}.
\newblock
\urldef\tempurl%
\url{https://doi.org/10.1126/science.aax2342}
\showDOI{\tempurl}
\showeprint{https://www.science.org/doi/pdf/10.1126/science.aax2342}


\bibitem[O'Neil(2017)]%
        {oneil2017weapons}
\bibfield{author}{\bibinfo{person}{Cathy O'Neil}.}
  \bibinfo{year}{2017}\natexlab{}.
\newblock \bibinfo{booktitle}{\emph{Weapons of math destruction: How big data
  increases inequality and threatens democracy}}.
\newblock \bibinfo{publisher}{Crown}, \bibinfo{address}{New York, NY}.
\newblock


\bibitem[Parker(2024)]%
        {wsjRealPage}
\bibfield{author}{\bibinfo{person}{Will Parker}.}
  \bibinfo{year}{2024}\natexlab{}.
\newblock \bibinfo{title}{Alleged Rent-Fixing of Apartments Nationwide Draws
  More Legal Scrutiny}.
\newblock
\newblock


\bibitem[Peng and Garg(2023)]%
        {peng2023monoculture}
\bibfield{author}{\bibinfo{person}{Kenny Peng} {and} \bibinfo{person}{Nikhil
  Garg}.} \bibinfo{year}{2023}\natexlab{}.
\newblock \bibinfo{title}{Monoculture in matching markets}.
\newblock
\newblock


\bibitem[Pleiss et~al\mbox{.}(2017)]%
        {pleiss2017calibration}
\bibfield{author}{\bibinfo{person}{Geoff Pleiss}, \bibinfo{person}{Manish
  Raghavan}, \bibinfo{person}{Felix Wu}, \bibinfo{person}{Jon Kleinberg}, {and}
  \bibinfo{person}{Kilian~Q Weinberger}.} \bibinfo{year}{2017}\natexlab{}.
\newblock \bibinfo{title}{On Fairness and Calibration}.
\newblock
\newblock


\bibitem[Raghavan et~al\mbox{.}(2020)]%
        {raghavan2020hiring}
\bibfield{author}{\bibinfo{person}{Manish Raghavan}, \bibinfo{person}{Solon
  Barocas}, \bibinfo{person}{Jon Kleinberg}, {and} \bibinfo{person}{Karen
  Levy}.} \bibinfo{year}{2020}\natexlab{}.
\newblock \showarticletitle{Mitigating Bias in Algorithmic Hiring: Evaluating
  Claims and Practices}. In \bibinfo{booktitle}{\emph{Proceedings of the 2020
  Conference on Fairness, Accountability, and Transparency}} (Barcelona, Spain)
  \emph{(\bibinfo{series}{FAT* '20})}. \bibinfo{publisher}{Association for
  Computing Machinery}, \bibinfo{address}{New York, NY, USA},
  \bibinfo{pages}{469–481}.
\newblock
\showISBNx{9781450369367}
\urldef\tempurl%
\url{https://doi.org/10.1145/3351095.3372828}
\showDOI{\tempurl}


\bibitem[Rawls(2004)]%
        {rawls2004theory}
\bibfield{author}{\bibinfo{person}{John Rawls}.}
  \bibinfo{year}{2004}\natexlab{}.
\newblock \bibinfo{booktitle}{\emph{A theory of justice}}.
\newblock \bibinfo{publisher}{Routledge}, \bibinfo{address}{England, UK}.
  229--234 pages.
\newblock


\bibitem[Reuters(2018)]%
        {amazonBias}
\bibfield{author}{\bibinfo{person}{Reuters}.} \bibinfo{year}{2018}\natexlab{}.
\newblock \bibinfo{title}{Amazon scraps secret AI recruiting tool that showed
  bias against women}.
\newblock
\newblock


\bibitem[Review(2021)]%
        {linkedinBias}
\bibfield{author}{\bibinfo{person}{MIT~Technology Review}.}
  \bibinfo{year}{2021}\natexlab{}.
\newblock \bibinfo{title}{LinkedIn’s job-matching AI was biased. The
  company’s solution? More AI.}
\newblock
  \bibinfo{howpublished}{\url{https://www.technologyreview.com/2021/06/23/1026825/linkedin-ai-bias-ziprecruiter-monster-artificial-intelligence/}}.
\newblock


\bibitem[Selbst et~al\mbox{.}(2019)]%
        {selbst2019fairness}
\bibfield{author}{\bibinfo{person}{Andrew~D Selbst}, \bibinfo{person}{Danah
  Boyd}, \bibinfo{person}{Sorelle~A Friedler}, \bibinfo{person}{Suresh
  Venkatasubramanian}, {and} \bibinfo{person}{Janet Vertesi}.}
  \bibinfo{year}{2019}\natexlab{}.
\newblock \showarticletitle{Fairness and abstraction in sociotechnical
  systems}. In \bibinfo{booktitle}{\emph{Proceedings of the conference on
  fairness, accountability, and transparency}}. \bibinfo{publisher}{ACM},
  \bibinfo{address}{Atlanta, GA}, \bibinfo{pages}{59--68}.
\newblock


\bibitem[Sen(1980)]%
        {SenEqualityOfWhat}
\bibfield{author}{\bibinfo{person}{Amartya Sen}.}
  \bibinfo{year}{1980}\natexlab{}.
\newblock \bibinfo{title}{Equality of What?}
\newblock
\newblock


\bibitem[Singh et~al\mbox{.}(2021)]%
        {singh2021fairness}
\bibfield{author}{\bibinfo{person}{Ashudeep Singh}, \bibinfo{person}{David
  Kempe}, {and} \bibinfo{person}{Thorsten Joachims}.}
  \bibinfo{year}{2021}\natexlab{}.
\newblock \showarticletitle{Fairness in ranking under uncertainty}. In
  \bibinfo{booktitle}{\emph{Advances in Neural Information Processing
  Systems}}, Vol.~\bibinfo{volume}{34}. \bibinfo{publisher}{Springer},
  \bibinfo{address}{Virtual}, \bibinfo{pages}{11896--11908}.
\newblock


\bibitem[Toups et~al\mbox{.}(2023)]%
        {toups2023ecosystem}
\bibfield{author}{\bibinfo{person}{Connor Toups}, \bibinfo{person}{Rishi
  Bommasani}, \bibinfo{person}{Kathleen Creel}, \bibinfo{person}{Sarah Bana},
  \bibinfo{person}{Dan Jurafsky}, {and} \bibinfo{person}{Percy~S Liang}.}
  \bibinfo{year}{2023}\natexlab{}.
\newblock \showarticletitle{Ecosystem-level Analysis of Deployed Machine
  Learning Reveals Homogeneous Outcomes}. In \bibinfo{booktitle}{\emph{Advances
  in Neural Information Processing Systems}},
  \bibfield{editor}{\bibinfo{person}{A.~Oh}, \bibinfo{person}{T.~Naumann},
  \bibinfo{person}{A.~Globerson}, \bibinfo{person}{K.~Saenko},
  \bibinfo{person}{M.~Hardt}, {and} \bibinfo{person}{S.~Levine}} (Eds.),
  Vol.~\bibinfo{volume}{36}. \bibinfo{publisher}{Curran Associates, Inc.},
  \bibinfo{address}{New Orleans, LA}, \bibinfo{pages}{51178--51201}.
\newblock
\urldef\tempurl%
\url{https://proceedings.neurips.cc/paper_files/paper/2023/file/a0b1082fc7823c4c68abcab4fa850e9c-Paper-Conference.pdf}
\showURL{%
\tempurl}


\bibitem[Venkatasubramanian and Alfano(2020)]%
        {venkatasubramanian2020recourse}
\bibfield{author}{\bibinfo{person}{Suresh Venkatasubramanian} {and}
  \bibinfo{person}{Mark Alfano}.} \bibinfo{year}{2020}\natexlab{}.
\newblock \showarticletitle{The philosophical basis of algorithmic recourse}.
  In \bibinfo{booktitle}{\emph{Proceedings of the 2020 conference on fairness,
  accountability, and transparency}}. \bibinfo{publisher}{ACM},
  \bibinfo{address}{Barcelona, Spain}, \bibinfo{pages}{284--293}.
\newblock


\bibitem[Waldman and Martin(2022)]%
        {Waldman2022}
\bibfield{author}{\bibinfo{person}{Ari Waldman} {and} \bibinfo{person}{Kirsten
  Martin}.} \bibinfo{year}{2022}\natexlab{}.
\newblock \showarticletitle{Governing algorithmic decisions: The role of
  decision importance and governance on perceived legitimacy of algorithmic
  decisions}.
\newblock \bibinfo{journal}{\emph{Big Data \& Society}} \bibinfo{volume}{9},
  \bibinfo{number}{1} (\bibinfo{date}{Jan.} \bibinfo{year}{2022}),
  \bibinfo{pages}{205395172211004}.
\newblock
\urldef\tempurl%
\url{https://doi.org/10.1177/20539517221100449}
\showDOI{\tempurl}


\bibitem[Wang et~al\mbox{.}(2024)]%
        {Wang2023}
\bibfield{author}{\bibinfo{person}{Angelina Wang}, \bibinfo{person}{Sayash
  Kapoor}, \bibinfo{person}{Solon Barocas}, {and} \bibinfo{person}{Arvind
  Narayanan}.} \bibinfo{year}{2024}\natexlab{}.
\newblock \showarticletitle{Against predictive optimization: On the legitimacy
  of decision-making algorithms that optimize predictive accuracy}.
\newblock \bibinfo{journal}{\emph{ACM Journal on Responsible Computing}}
  \bibinfo{volume}{1}, \bibinfo{number}{1} (\bibinfo{year}{2024}),
  \bibinfo{pages}{1--45}.
\newblock


\bibitem[Williams(1962)]%
        {williams1962warrior}
\bibfield{author}{\bibinfo{person}{Bernard Williams}.}
  \bibinfo{year}{1962}\natexlab{}.
\newblock \showarticletitle{The Idea of Equality}.
\newblock In \bibinfo{booktitle}{\emph{Philosophy, Politics and Society}},
  \bibfield{editor}{\bibinfo{person}{Peter Laslett} {and}
  \bibinfo{person}{Walter Runciman}} (Eds.). \bibinfo{publisher}{Blackwell},
  \bibinfo{address}{Oxford}, \bibinfo{pages}{112--117}.
\newblock


\bibitem[Zliobaite(2015)]%
        {zliobaite2015relation}
\bibfield{author}{\bibinfo{person}{Indre Zliobaite}.}
  \bibinfo{year}{2015}\natexlab{}.
\newblock \bibinfo{title}{On the relation between accuracy and fairness in
  binary classification}.
\newblock
\newblock


\end{thebibliography}

\end{document}